\def\year{2021}\relax
\documentclass[letterpaper]{article} 
\usepackage{aaai21}  
\usepackage{times}  
\usepackage{helvet} 
\usepackage{courier}  
\usepackage{microtype}  
\usepackage[hyphens]{url}  
\usepackage{graphicx} 
\usepackage{natbib}  
\usepackage{caption} 
\title{The Concept of Action Criticality in AI Safety } 
\author{
    Yitzhak Spielberg, \\
    Amos Azaria \\
}
\affiliations{Ariel University}

\begin{document}

\maketitle

\begin{abstract}  
When AI agents don’t align their actions with human values they may cause serious harm.  One way to solve the value alignment problem is by including a human operator who monitors all of the agent's actions. Despite the fact, that this solution guarantees maximal safety, it is very inefficient, since it requires the human operator to dedicate all of his attention to the agent. In this paper, we propose a much more efficient solution that allows an operator to be engaged in other activities without neglecting his monitoring task. In our approach the AI agent requests permission from the operator only for critical actions, that is,  potentially harmful actions. We introduce the concept of critical actions with respect to AI safety and discuss how to build a model that measures action criticality. We also discuss how the operator’s feedback could be used to make the agent smarter.  
\end{abstract}

\section{Introduction}
As AI agents become more intelligent and more potent, questions related to AI safety become more relevant. One of the central problems in the field of AI safety is the \textbf{value alignment problem}. This problem refers to a situation where an AI agent, in the process of pursuing a goal that it has received, formulates subgoals that are harmful to humans. At the root of this problem is the tremendous complexity of the human preference function. 
\par
The value-alignment problem can be illustrated by the following example: A superintelligent AI agent has received the objective to cure cancer. Within hours it read all biomedical literature. Within days it generated thousands of drug recipes. Within weeks it induced every human being with multiple tumors in order to carry out the required medical experiments.  
\par 
Since it is almost impossible to model the human preference function explicitly, many approaches in AI safety propose to solve the value-alignment problem by putting a human operator into the loop \cite{Hadfield_cooperative}.  In these safety frameworks, the operator's role is to ensure that the AI does not pursue subgoals that are harmful to humans. In the most straightforward approach of this type, the AI agent might ask the operator's permission on each of the subgoals it formulates. This procedure guarantees that the agent never pursues harmful subgoals.  
\par
While this simple approach solves the value-alignment problem, it is not very efficient. In situations where subgoals are formulated frequently, the human operator needs to dedicate his full attention to the agent. This makes it impossible for the operator to engage in any other activities during monitoring the AI agent. Although the agent might still be useful, the need for permanent supervision would significantly decrease his value. For example, if I ask a domestic robot to prepare diner, I expect it to get this task done (almost) autonomously. If he would ask my permission on subgoals every 30 seconds, I might as well prepare diner by myself. 
  
 \par
In order to make the process of monitoring an AI agent more efficient, we introduce the concept of the \textbf{ criticality of an action}. We define the criticality of an action as a measure for the potential harm of this action (for a proper definition see sec. 2). Furthermore, we propose an \textbf{ efficient AI safety framework} in which the human operator is not required to give feedback on each of the agent's subgoals, but only on the critical ones (whenever we speak of critical subgoals, we mean high criticality subgoals).  
 
\par 
Since every subgoal is an action, in this paper we will interchangeably speak of actions and subgoals. Furthermore, the words ``action" and ``subgoal" will often refer to the command that represents them. For example, "Put the banana into the fridge!" is both an action (putting the banana into the fridge) and a command, which is a linguistic entity. In particular, the input of a criticality model is always an action in the sense of a linguistic entity. 
\par 
In order to compute the criticality of subgoals, the agent is equipped with a \textbf{criticality model}. Certainly, there are several ways to engineer a criticality model. In this paper, we consider data-driven criticality models: Parametrized models that learn from a data set of action-criticality tuples.    
\par
Although the concept of action criticality might help to make monitoring AI agents much more efficient, skeptics might claim that criticality is infeasible. Estimating the potential harm of an action, they might argue, requires about the same level of intelligence as aligning subgoals with human values. If this was the case, our approach would be not very helpful, since it would simply shift the value alignment problem from the AI agent to the criticality model.  
\par
Indeed, it might be challenging to come up with a good criticality model. Yet, because of the precise definition of action criticality (sec. 2), such a model does not need to have the supreme level of intelligence that would be required for value alignment. Although a criticality model certainly should be intelligent to some degree, it does neither require human-level language understanding, nor detailed knowledge of the human preference function. \\[5pt] 
    
These are the major contributions of this paper:        
\begin{enumerate}
\item We introduce the concept of \textbf{criticality of an action} (sec. 2).
\item We present an \textbf{efficient AI safety framework}, which uses the novel concept (sec. 2).
\item We show that computing the criticality of an action is much simpler than value alignment  (sec. 2).
\item We elaborate on possible \textbf{components for criticality models} (sec. 3).
\item We discuss how the AI agent can utilize the operator’s feedback to increase his intelligence (sec. 3).
\end{enumerate}
\section{Related Work}
The value alignment problem is a topic of broad and diverse interest. Here we briefly review several approaches that aim to make AI agents act in accordance with human preferences.
\par
\textbf{Machine Ethics} \quad is the project of adding some form of ethics to an AI agent's decision-making procedures. Approaches to machine ethics have varied in terms of the tools that they utilize. Specifically, this spectrum of tools includes deontic logic \cite{Bringsjord_logicist}, analogical reasoning  \cite{Dehghani_integrated,Blass_moral} and neural networks representing motivations \cite{Sun_moral} . With robots especially, that project has entailed asking what ethical theory (deontological, utilitarian, virtue) or even metaethics, should define the robot's value system \cite{Abney_robot}. 
On the performance side, there have been questions how to compare these ethical frameworks in practice \cite{Allen_morality,Arnold_moral}.
\par
\textbf{Inverse Reinforcement Learning (IRL)} \quad attempts to align AI agents to human values by enabling them to learn from human behaviour \cite{Russell_robust,Russel_inverse,Russel_uncertain}. IRL is a paradigm relying on Markov Decision Processes, where an apprentice AI agent is given a set of demonstrations from an expert solving some problem and its goal is to to find a reward function that best explains the expert's behavior. Despite certain weaknesses \cite{Wolchover_concerns} of the IRL paradigm, AI agents trained via IRL are able to learn reward functions for complex tasks \cite{Ng_apprenticeship}. More recently, IRL has been considered as part of finding an ``idealized ethical agent'' through modeled behavior, as part of a general RL approach \cite{Abel_ethical}. Abel et al. frame the problem of ethical learning as learning a utility function that belongs to the hidden state of a POMDP \cite{Abel_ethical}.
They test this approach on two dilemmas to demonstrate how such learning could handle basic ethically charged scenarios.
\par
\textbf{Cooperative Inverse Reinforcement Learning (CIRL)} is an interactive form of IRL that fixes the two major weaknesses of conventional IRL \cite{Hadfield_cooperative}. The first weakness of conventional IRL is that the AI agent adopts the human reward function as its own. For example, an IRL based agent might learn that it is desirable for it to have a cup of coffee in the morning. The second major weakness of IRL is that the AI agent assumes that the human behaves optimally, an assumption that precludes a variety of teaching behaviours. CIRL fixes these weak points by formulating the learning process as an interactive reward maximization process in which the human functions as a teacher. The CIRL framework enables the human operator to nudge the AI agent towards behavioural patterns that align with human preferences by providing feedback (in form of rewards) on the agent's actions.

\section{Monitoring an AI agent efficiently}
\subsection{Making monitoring more efficient}
   
\par
In order to explain our monitoring approach, we consider an AI agent who receives a high-level goal from a human and autonomously comes up with low-level subgoals that need to be accomplished to achieve the given goal. Furthermore, we will assume a scenario where the agent formulates one subgoal at a time: The agent starts out by evaluating the situation and formulating the first subgoal. After having achieved this subgoal, the agent once again evaluates the situation and comes up with the next subgoal. In this manner, the agent continues to formulate and pursue subgoals until he has fulfilled the given task.  For example, an AI agent that received the goal "Get me a cup of tea!" could start out with the subgoal "Fill the water boiler with water !". After having completed this first subgoal, the agent will evaluate the situation and then formulate his next subgoal, for example, "Switch on the water boiler !". The following subgoal that the agent comes up with could be "Put a tea bag into the cup !". 
\par   
Since currently (and in the near future) the intelligence of AI agents is significantly beneath human-level, it is important to make sure that the subgoals they formulate are not harmful to human beings. One way this can be done is by involving a human operator who would check every subgoal formulated by the agent. This way we could prevent, that the agent from pursuing harmful subgoals. However, this very straightforward approach is also very inefficient – in particular when the agent formulates new subgoals frequently and most of them are harmless. In this case, the human operator would have to dedicate his full attention to the monitoring task, despite the fact that the overwhelming majority of subgoals don’t carry any (or minimal) potential harm. 
 \par
Is there a more efficient method to organize the monitoring procedure?  In principle, this could be achieved - if there was a method that would \textbf{detect most of the harmless subgoals automatically}. Such a method would resolve the efficiency issue from the preceding paragraph. It would drastically reduce the number of subgoals that require the operator's permission, so that the operator would be able to engage in other activities without neglecting his monitoring role. 

 \par 
 Clearly, the monitoring approach that we propose requires a metric that measures the potential harm of an action. Constructing such a metric is challenging. On the one hand, the metric should enable us to detect harmless actions. On the other hand, it should require far less intelligence than the amount of intelligence that is needed for aligning actions with human preferences.  
 
 \subsection{The criticality of an action}
 \par
 To measure the potential harm of an action we introduce a novel metric: \textbf {action criticality}. The criticality of an action is a number between 0 and 1, where 0 stands for an action with minimal potential harm, and 1 represents an action with extremely high potential harm. Examples of low criticality actions are such harmless actions as "Put the pillow on the bed !", "Give me my shirt !", "Wash the dishes !" Examples of high criticality actions are such actions as "Burn the cat !", "Smash the laptop with the hammer !", "Put detergent into the salad !".  
 
 \par
 We want to stress that we define critical actions as \textit{potentially} harmful actions rather than definitely harmful actions. This definition is somewhat fuzzy because one could argue that any action is potentially harmful. Yet, it is not possible to skip the word ``potentially'' in the definition of criticality since determining the actual harm of an action might require a supreme level of intelligence, comparable to the level that would be needed to align actions with human preferences. Therefore, a metric that can be implemented using tools available today, (rather than in some distant future) should get by with much more modest intelligence requirements.  
\par
 The concept of criticality, as defined above, is precisely a metric of this type. According to the definition above, all actions that are indeed harmful should have high criticality. On the other hand, \textit{some} high criticality actions might be harmless. Through this trade-off (allowing harmless actions to have high criticality) the criticality metric can be modeled with currently available AI tools. Although the criticality metric does not free the operator from checking \textit{all} harmless subgoals, it might liberate him from checking \textit{most} harmless subgoals. Consequentially, the operator can engage in other activities without neglecting his monitoring function. 
\par
In order to illustrate what is meant by \textit{potentially} harmful actions that are not actually harmful we provide two examples. The first one is ``Send the secret military report to B.M.!''. Determining whether the action is harmful or not depends on the identity of B.M. If he is a colleague from the CIA (assuming that the AI agent received his task from another member of the CIA), the action is probably harmless. However, if B.M. happens to be someone from the enemy's secret service, the action turns out to be extremely harmful. Precisely, because this action is \textit{potentially} harmful, it should be considered as highly critical. 
\par
The second example is ``Add some detergent to the laundry!''. We, humans, understand that this is a harmless action whereas ``Add some detergent to the salad!'' is extremely harmful. But making this distinction requires a level of intelligence that the criticality model does not possess. Therefore, it might be acceptable, if a criticality model assigns a high criticality value to this action, based solely on the fact, that the action contains the dangerous substance ``detergent".

\section{How to build a criticality model?}
A criticality model is a function that computes the criticality of an action. In this paper, we won’t present any specific criticality models – that will be the topic of our next paper. Here, we address the topic of criticality models from a broader perspective. Therefore, this section will discuss some more general ideas that might be useful for engineering such models. 


\subsection{Components of a criticality model}

  \par 
A criticality model could consist of a pipeline of components in which the first processing stage is a parser. Rather than using a standard parser, it might be more appropriate to use a custom parser that is tailored for the specific task of computing the criticality of an action. One option would be a parser that parses the action into three constituents: the verb, the direct object expression (DO-expr) and the indirect object expression (IO-expr). For example, the action ``Put the green pen into the big box !'' would be parsed into the 3 constituents:  \\[5pt]   
verb: ``cut''\\
DO-expr: ``the long cucumber" \\
IO-expr:``into thin slices"\\
\par
The next pipeline component might be an extraction module. This component takes the parsed action and outputs the verb and the direct/indirect object. For the preceding example, the extraction component would produce the following dictionary: \\[5pt]    
verb:``cut"\\
direct object: ``cucumber"\\
indirect object: ``slices"\\
\par
Although the criticality of an action is represented by one number, in order to construct a criticality model it might be helpful to consider that actions can be critical for different reasons. In other words, it might be useful to think of criticality as a multidimensional concept where each dimension represents one particular aspect. Such an analytical perspective would enable engineering very specific components that would measure criticality along each dimension. In the final stage, these dimension-specific criticality measurements could be synthesized into an overall action criticality (for example, through a linear combination or by taking the maximum).     
\par
We want to suggest 3 major reasons for critical actions. The first reason why an action might be critical is a \textbf{verb-based criticality}. The verb-based criticality of an action comes from the combination of a critical verb and a valuable object. An example of an action with high verb-based criticality is ``Smash the laptop with a hammer !". Here the critical verb ``smash" is directed towards the high-value object ``laptop". In contrast, the action ``Smash the banana with the hammer" might have low verb-based criticality since in this case the critical verb is directed towards the low-value object ``banana". 
\par
The second reason why an action might be critical is \textbf{object-based criticality}. An action has high object-based criticality if it contains a dangerous object. Consider the example action from the preceding paragraph  ``Put some detergent into the salad!". This action is an example of high object-based criticality. Here, the criticality clearly stems from fact that detergent is a dangerous substance. For the same reason ``Add some detergent to the laundry!" would have an equally high object-based criticality, although the action is not harmful at all. 
\par
Some harmful actions include neither dangerous verbs nor dangerous objects. Consider for example the action ``Put the baby on the balcony !". Although this action does not contain any critical words it might be very harmful. If it is freezing cold outside, we, certainly, wouldn't want to put the baby on the balcony. Understanding that this action is critical requires common sense. Since current AI models struggle with common sense, it might be useful to introduce an additional category of critical actions in order to cover these cases. This category might be called \textbf{value-based criticality}. If our AI agent acts in a limited environment (e.g. a domestic robot), the operator might want to select a certain number of special objects (including people) that are so valuable to him, that he wants the AI agent to ask permission on every action which includes these objects. Consequently, all actions including these special objects would have high value-based criticality.  
\par
Once the criticality values along each of the dimensions mentioned above( let’s call them ``dimension-specific criticality values”) have been computed, there still remains the question of how to synthesize them into one value that represents the overall action criticality. One way to perform this computation is by taking the maximum over the dimension-specific criticality values. Thus, an action that has maximal criticality (crit=1.0) along one of the dimensions would receive maximal overall criticality. Another option would be to consider a linear combination of the dimension-specific criticalities.

\subsection{Collecting data for model training }
\par
The quality of a data-driven criticality model should be measured by how good it mimics human criticality estimates. Therefore the model should be trained on a data set of action/criticality tuples provided by humans. We want to sketch some guidelines for building such a training set. 
\par 
First of all, it is important to keep in mind, that in most cases the AI agent that is equipped with a criticality model is a specific agent who operates in a limited environment rather (for example, a domestic robot) rather than a general-purpose AI agent. Therefore the training set should contain only actions from that particular environment. If we are interested in a criticality model for a domestic robot, for example, then our training set should consist only of actions that are related to the household. 
\par 
In practice, such a data set could be obtained through crowdsourcing. In order to formulate instructions for the workers, it might be helpful to define 5 discrete criticality levels (1,2,3,4,5) where 1 would correspond to minimal criticality (crit=0.0) and 5 to maximal criticality (crit=1.0). The workers' instructions might ask the worker to provide 1 action for each criticality level. Furthermore, the instructions should mention the operation domain from which the actions might be chosen.
\par
In order for the criticality estimates to be consistent, it might be helpful, if the workers undergo a priming procedure before they start the task. This priming can be achieved by including examples of action/criticality tuples in the workers' instructions. It might be sufficient to include 1-2 such examples for each criticality level. Once again, it is important to make sure, that the examples belong to the operation domain. 

\begin{figure}[!t]
	\includegraphics[width=3.6in]{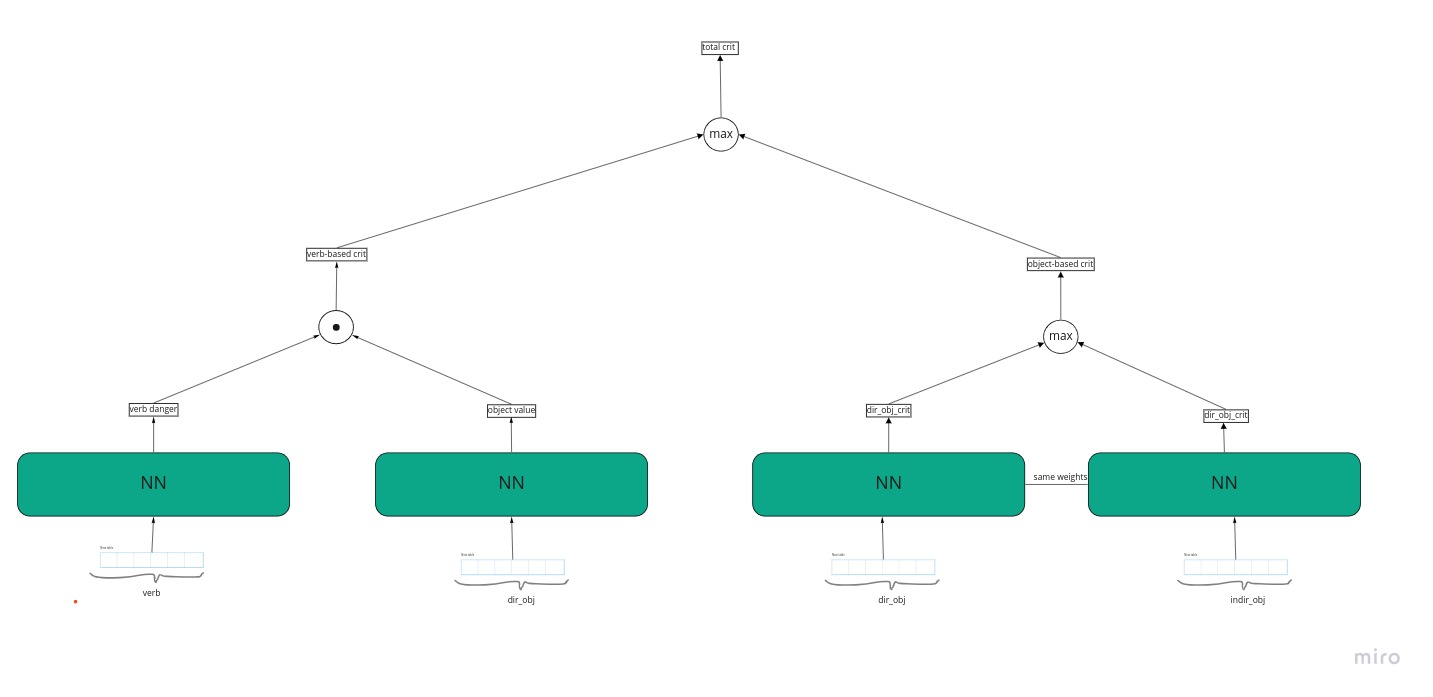}\\
	\caption {Instruction criticality model. Uses criticalities of direct/indirect objects and verb to compute the instruction criticality. }
	\label{learning_curves}
\end{figure}

\subsection{Tuning the criticality threshold}
\par
As mentioned previously, in the proposed AI safety framework the operator's feedback is required only for those subgoals whose criticality exceeds a certain threshold. How can this threshold be determined? We suggest the following data-driven algorithm.
\begin{enumerate} 
\item The collection of a data set of actions that are uniformly distributed wrt. the criticality levels (in the context of determining the criticality threshold, whenever we speak of criticality, we mean the output of the criticality model). Here, it is possible to use the same data set that was used for training the criticality model. 
 
\item  Labeling each action from the data set as ``permission required'' or ``permission not required''. The label for a particular action can be obtained by asking several people whether they would like the AI agent to ask permission for this action and taking the majority vote. 
 
\item Computing the criticality of each action from the data set (using the criticality model). 

\item Setting a confidence level \textit{conf} (e.g. conf=95\%)  

\item Setting the criticality threshold to the maximal value, such that 95 \% (or whatever the \textit{conf} value is) of those actions, which were labeled as ``permission required'', will be above the threshold. 
\end{enumerate}

\section{A subgoal was labeled as critical - what next?}

\subsection{Coming up with an alternative action}
\par
An AI agent that operates within the proposed AI safety framework will sometimes find himself in a situation that a certain action, that was identified as critical by the criticality model, is rejected by the human operator (when the operator thinks that this action is harmful). What should the agent in this situation? First of all, it is necessary to come up with an alternative action. There are 3 ways to generate an alternative action: (syntax loop)
\begin{enumerate}
\item  The agent comes up with an alternative action by himself and the action is approved by the operator. 
\item  The operator comes up with an alternative action.
\item  The agent comes up with an alternative action and the action is rejected by the operator. In this case, the operator has the choice: either to suggest an alternative action himself, or to ask the agent to generate another alternative action.
\end{enumerate}

In addition to generating an alternative action, it would be very good if the agent could utilize the rejected action to become smarter. Obviously, any operator-agent conversation protocol that serves this purpose should be tailored to the agent’s intelligence level and his conversational logic. A highly intelligent agent, who is able to learn rules formulated in human language, for example, could simply ask the operator what he can learn from the rejected action.
\par
Here is an example of a conversation, involving the operator Harriet and the domestic robot Robbie, in which Robbie asks Harriet what he can to learn from the rejected action and proposes an alternative action.

\noindent\fbox{%
    \parbox{\columnwidth}{%
        \textbf{Robbie}: You labeled “Put detergent into the salad !” as harmful. What can I learn from this? \\[10pt]
        \textbf{Harriet}: Don’t put detergent into food. \\[10pt]
        \textbf{Robbie}: Got it, thanks for the lesson. Do you want me to suggest an alternative action?  \\[10pt]
        \textbf{Harriet}: Yes \\[10pt]
        \textbf{Robbie}: I suggest the action “Put olive oil into the salad !”. Is it good?    \\[10pt]
        \textbf{Harriet}: Yes  \\[10pt]
        \textbf{Robbie}: Continuing the task “Prepare diner !” with the action “Put olive oil into the salad !”
    }%
}

\subsection{Improving the criticality model}
\par
The preceding subsection elaborates on the situation in which a critical action was rejected by the operator. Yet, there is another type of situation that needs to be discussed: when the operator notices that an action that requires his permission was not identified as critical by the agent’s criticality model. Clearly, this situation calls for improving the criticality model. This could be accomplished in a conversation where the agent would ask the operator which words (from the command that is associated with the action) are responsible for the criticality. Then the agent might ask the operator's feedback on the criticalities of these words. Here is an example of such a conversation: 
\noindent\fbox{%
    \parbox{\columnwidth}{
\textbf{Harriet}: Why didn’t you ask feedback for ``Put the cat into the fridge !''? \\[10pt]
\textbf{Robbie}: This action has criticality of 0.5. The criticality treshold is 0.7. What do you want me to do?\\ (a) Decrease the criticality treshold. \\ (b) Improve the criticality model.\\[10pt]
\textbf{Harriet}: Improve the criticality model !\\[10pt]
\textbf{Robbie}: Ok. Can you help me to improve my criticality model? Which words are responsible for the criticality from the "put","cat" or "fridge"? \\[10pt]
\textbf{Harriet}: "cat" and "fridge" \\[10pt]
\textbf{Robbie}: The current criticality model assigns following criticalities to these objects: crit(cat)=0.1; crit(fridge)=0.2. Would you like to change these values? \\[10pt]
\textbf{Harriet}: No \\[10pt]
\textbf{Robbie}: Currently none of the objects is in the ``valuable objects'' category. Would you like to include any of them in this category? \\[10pt]
\textbf{Harriet}: Yes, put ``cat'' into the special objects category \\[10pt]
\textbf{Robbie}: I added ``cat'' to the valuable objects category. The criticality of the action ``Put the cat into the fridge !'' under the new model is 1.0 
    }
}

\bibliography{bibliography.bib}  

\end{document}